# Optical Detection of the Hidden Nuclear Engine in NGC 4258

Belinda J. Wilkes[1]

belinda@cfa.harvard

Gary D. Schmidt[2]

gschmidt@as.arizona.edu

Paul S. Smith[2]

psmith@as.arizona.edu

Smita Mathur[1]

mathur@cfa.harvard.edu

Kim K. McLeod[1]

kmcleod@cfa.harvard.edu

[1]Harvard-Smithsonian Center for Astrophysics, 60 Garden St., Cambridge, MA 02138

[2]Steward Observatory, University of Arizona, Tucson, Arizona 85721

## ABSTRACT

The sub-parsec masing disk recently found to be orbiting a central mass of $\sim 3.6 \times 10^7$ M$_\odot$ in the Seyfert/LINER galaxy NGC 4258 (Miyoshi *et al.* 1995) provides the most compelling evidence to date for the existence of a massive black hole in the nucleus of a galaxy. The disk is oriented nearly edge-on and the X-ray spectrum is heavily absorbed. Therefore, in this galaxy, the optical emission-line spectrum generally exhibited by an active galactic nucleus is perhaps best sought using polarized light: probing for light scattered off material surrounding the central source. New polarimetry of NGC 4258 has uncovered a compact polarized nucleus whose spectrum consists of a faint blue continuum similar to those of unobscured quasars ($F_\nu \propto \nu^{-1.1}$), plus broadened ($\sim$1000 km s$^{-1}$) emission lines. The lines are strongly linearly polarized (5 − 10%) at a position angle (85°±2°) coincident with the plane of the maser disk. This result provides substantiating evidence for a weakly active central engine in NGC 4258 and for the existence of obscuring, orbiting tori which impart many of the perceived distinctions between various types of active galaxy.

*Subject headings:* polarization – galaxies: individual: NGC 4258 – galaxies: active – galaxies: nuclei

## 1. Introduction

NGC 4258 (M106) is a giant SABbc galaxy ($v = 472\pm4$ km s$^{-1}$, Cecil, Wilson, & Tully 1992) at a distance of $\sim$ 7 Mpc (Tully 1988). Its morphology is complex, with inner and outer spiral arms (Courtés *et al.* 1993), a dust lane and an optical/radio jet at a position angle ($\sim 150°$; Cecil *et al.* 1992) close to that of the galaxy major axis. It has a LINER/Seyfert 2 emission-line spectrum whose classification has been long debated. Early reports of the presence of broad wings to H$\alpha$ led to a Seyfert 1.9 classification (Stauffer 1982; Filipenko & Sargent 1985). These results have since been called into question by Stüwe, Schulz, & Hühnermann (1992), who argue that fitting the emission lines with Lorentzian rather than Gaussian profiles avoids the need for a broad H$\alpha$ component. The strengths of the central (<5") emission lines suggest the presence of a weak, central ionizing continuum (Q$\sim 10^{52}$ s$^{-1}$, $F_\nu \propto \nu^{-1.2}$; Stüwe *et al.* 1992) significantly stronger than the observed UV continuum (Ellis, Gondhalekar, & Efstathiou 1982).

The most compelling evidence for a past/present active nucleus derives from radio measurements. High spatial resolution observations of the well-known, high-velocity H$_2$O maser emission have revealed a thin, edge-on disk (height/radius$\leq$ 0.02, $i = 83°$) with inner and outer radii of 0.13 and 0.25 pc, respectively (Miyoshi *et al.* 1995). The rotational velocity is consistent with Keplerian motion and ranges from 700–1000 km s$^{-1}$ across the disk, implying a binding mass of $\sim 1.4 \times 10^7$ $M_\odot$. This result is highly suggestive of the presence of a central black hole surrounded by a disk of material sufficiently thick to sustain masing molecules. In addition, a compact radio source coincident with the nucleus has been detected between 2 cm and 20 cm (Miyoshi *et al.* 1995; Hummel, Kraus, & Lesch 1989).

X-ray observations of NGC 4258 (Makishima *et al.* 1994) detected an unresolved ($\leq 1'$), strongly absorbed source coincident with the optical nucleus. The authors suggest that this emission originates in a low-luminosity ($\sim 4 \times 10^{40}$ ergs s$^{-1}$) active nucleus which is heavily obscured along our line of sight (equivalent $N_H \sim 10^{23}$ cm$^{-2}$).

The presence of a disk coupled with strong obscuration of the nuclear regions makes NGC 4258 a prime candidate in which to search for hidden broad emission lines, visible in polarized light alone, as are often seen in otherwise narrow emission-line objects (Miller 1994). In this paper we report the detection of a polarized blue continuum and emission lines from the central regions of NGC 4258 which portray an active nucleus hidden from our direct view. This galaxy provides a direct link between the obscuring torus often inferred in an active galactic nucleus (AGN) and a disk of material orbiting a black hole of high mass.

## 2. Imaging and Spectropolarimetry of NGC 4258

Observations were made on several occasions between 1994 November 26 and 1995 June 20 using the CCD Spectropolarimeter described by (Schmidt, Stockman, & Smith 1992) at the 2.3 m telescope of Steward Observatory and the 1.9 m Perkins telescope of Lowell Observatory/Ohio State Univ. Spectropolarimetry utilized a slit of 3″ or 4″ width centered on the core of the optical image, and oriented either NS or EW. For the imaging observations, a plane mirror was substituted for the grating and a large entrance aperture provided polarization maps in broad wavelength bands over the central 50″×50″ region of the galaxy with 0″.5 square pixels.

A polarization map made in the $V$ photometric band (4900 − 6000 Å) is shown as Figure 1. Here, vector length is proportional to the local surface brightness of polarized light, while orientation indicates the electric vector position angle. Significant polarization ($p = 0.1 - 1\%$) due to selective extinction by aligned dust grains in the interstellar medium of NGC 4258 exists throughout the map. The degree of polarization of the nucleus itself is small, $p \sim 0.25\%$, but because of its brightness is the dominant source of polarized flux in the map. The PA measured in both $V$ and $R$ (not shown) is essentially EW – identical to the 86° PA of the maser disk on the sky (Miyoshi *et al.* 1995). The polarized nucleus is highly concentrated in both bands, with no detectable NS extension above the seeing profile (FWHM = 1".5) and a possible small EW extension of $\lesssim 3"$.

Spectropolarimetric results for a $3"\times 7"$ slit centered on the nucleus are depicted in Figure 2. These are our highest-resolution observations, covering the interval $\lambda\lambda 4580 - 7110$ at a resolution of 400 km s$^{-1}$ FWHM. The total flux spectrum (*top*) shows well-known LINER/Seyfert features: narrow ($\sim$500 km s$^{-1}$) emission lines of [OIII] $\lambda\lambda$4959, 5007, H$\alpha$, [NII] $\lambda\lambda$6548, 6583 and [SII] $\lambda\lambda$6716, 6731. In Stokes flux $q'\times F_\lambda$ (equivalent to polarized flux, *bottom*), these lines appear both broader ($\sim$1000 km s$^{-1}$) and in different relative strengths. In addition, H$\beta$ and [OI] $\lambda\lambda$6300, 6363 may be weakly present in polarized flux. Measurements of line flux, width, and polarization are provided in Table 1. Uncertainties in the fluxes and widths are estimated to be $< 20\%$ except where noted by a ":". The PAs of all lines are consistent, and the co-added value for all features, 85°±2°, matches the PA of the plane of the maser disk. The total H$\alpha$ flux measured in our $3"\times 7"$ extraction aperture, $4.5\times 10^{-14}$ ergs cm$^{-2}$ s$^{-1}$, is somewhat lower than the $7.41\times 10^{-14}$ ergs cm$^{-2}$ s$^{-1}$ found for a $4"$ aperture by Stauffer (1982).

The degree of polarization in the continuum rises smoothly to the blue from a value of $p = 0.18\%$ at 7000 Å to $p = 0.29\%$ at 4800 Å. The measured Stokes flux at 5500 Å, $3.2 \pm 0.3 \times 10^{-17}$ erg cm$^{-2}$ s$^{-1}$ Å$^{-1}$, is equivalent to a visual magnitude of 20.2. A power law provides an adequate fit to the spectral dependence over our observed range, $\lambda\lambda 4150 - 8000$, yielding $q'\times F_\nu \propto \nu^{-1.1\pm 0.2}$, similar to the nonthermal spectra of many unobscured AGN. The PA differs slightly from the emission-line value and possibly between the two orthogonal slit positions. Both discrepancies are probably the result of contributions by differently-polarized, off-nuclear regions of the galaxy which are contained within the extraction apertures.

## 3. A Hidden Active Nucleus seen in Scattered Light

The close agreement between the PAs of the apparent major axis of the maser disk and the emission-line and continuum polarization suggests that all three are related to a common geometry. Synchrotron emission is ruled out by the polarized emission lines, while the blue continuum and high line polarization ($p_{H\alpha} \sim 9\%$) argues strongly against extinction by aligned dust grains. Because the disk is oriented nearly edge-on ($i = 83°$, Miyoshi *et al.* 1995), a consistent explanation is that light emitted in the nucleus is able to escape the disk through axial holes and is subsequently scattered off particles situated above and/or below the disk. For near right-angle scattering, the polarization of the reflected light can be strong, with an electric vector perpendicular to the axis of the disk, or parallel to its apparent image on the sky. The phenomenon is identical to that which permits the detection of hidden broad emission-line regions in the nuclei of many Seyfert 2 galaxies (Antonucci & Miller 1985; Miller 1994).

In the optically-thin limit, the surface brightness of light scattered off material of characteristic dimension $l$ situated at a distance $r$ from a source of luminosity $L_\lambda$ and composed of particles with number density $N$ and scattering cross-section $\sigma_\lambda$ can be written[3]

$$B_\lambda \sim \frac{N\sigma_\lambda l L_\lambda}{16\pi^2 r^2} \ .$$

Since the polarimetric images appear marginally resolved, a reasonable value for $l$ and $r$ might be 1″ (35 pc). The Stokes flux measured at 5500 Å of $3.2 \times 10^{-17}$ erg cm$^{-2}$ s$^{-1}$ Å$^{-1}$ implies a polarized surface brightness of $p \times B_{5500} \sim 2 \times 10^{-17}$ erg cm$^{-2}$ s$^{-1}$ Å$^{-1}$ arcsec$^{-2}$ if we assume two identical 1″ diameter scattering clouds. The intrinsic degree of polarization of the scattered light is unknown, but must exceed 9% – the net polarization of H$\alpha$ – and be less than 100%. We take $p = 20\%$ to be representative. This leads to an estimate for the luminosity of the nuclear continuum source of $L_{5500} \sim 6 \times 10^{16}/N\sigma$ ergs s$^{-1}$ Å$^{-1}$. Scattering by electrons for example in HII regions surrounding the nucleus has the feature that the cross-section is wavelength-independent ($\sigma_e = 6.65 \times 10^{-25}$ cm$^2$), but places the added restriction that the resulting recombination radiation not exceed what is measured. For the assumed material, the observed (total) H$\alpha$ flux of $4.5 \times 10^{-14}$ ergs cm$^{-2}$ s$^{-1}$ requires $N_e \lesssim 30$ cm$^{-3}$, leading to a nuclear luminosity of $L_{5500} \gtrsim 3 \times 10^{39}$ ergs s$^{-1}$ Å$^{-1}$ ($\lambda L_{5500} \sim 10^{10}$ L$_\odot$). Material located closer to the nucleus would require a lower luminosity continuum source, or a lower particle density, to achieve the same scattered surface brightness.

---

[3] An error in the derivation of this formula by Dutil *et al.* (1995) led them to underestimate the surface brightness by a factor of $10^9$ and conclude that scattering is not viable.

Astrophysical dust grains are found in a variety of sizes and compositions, but in general have scattering cross-sections per unit mass far greater than that for electrons. For example, a "typical" interstellar grain from the mixture of Mathis, Rumpl, & Nordsieck (1977) has a radius of $\sim 0.05$ $\mu$m and scattering cross-section in the visible of $\sim 1 \times 10^{-12}$ cm$^2$ (White 1979). If these populate the above clouds with a typical gas-to-dust ratio of $\rho_H/\rho_D \sim 100$, the scattered surface brightness off grains would exceed that from electrons by nearly two orders of magnitude. The required nuclear illuminating source is commensurately fainter, $\lambda L_{5500} \sim 10^8$ L$_\odot$. It should be noted that the blackbody temperature for grains $\sim 35$ pc from the central source is less than 20 K, well below their sublimation temperature.

Using the scattering estimates above as a guide and assuming negligible extinction toward the scattering material, a luminosity $L_{5500} \sim 10^{37} - 10^{39}$ ergs s$^{-1}$ Å$^{-1}$ coupled with the absorption-corrected 2 – 10 keV flux of $6.4 \times 10^{-12}$ ergs cm$^{-2}$ s$^{-1}$ (Makishima et al. 1994) implies an intrinsic optical-to-X-ray spectral index for the nucleus of $1.0 < \alpha_{\rm ox} < 1.8$, spanning the normal range for AGN. The bolometric nuclear luminosity determined from the polarized flux is $\sim 10^{42-44}$ erg s$^{-1}$, comparable to that deduced from the emission lines (Stüwe et al. 1992) and consistent with the luminosity expected for an AGN with central

– 8 –

mass $\sim 10^7$ M$_\odot$ (Wandel & Mushotzky 1986). Figure 3 compares the total and polarized flux spectral energy distributions of NGC 4258 with that of a typical radio-quiet quasar (Elvis *et al.* 1994).

According to the line ratio criteria of Veilleux & Osterbrock (1987), the polarized flux spectrum of NGC 4258 would be classified among narrow-line AGNs, the single exception being the relative weakness of [SII] $\lambda\lambda$6716, 6731. The line width of $\sim$1000 km s$^{-1}$ is somewhat broader than most Seyfert 2 nuclei, but might be associated with motion in a disk whose outer regions are seen in the light of H$_2$O masers.

## 4. Conclusions

Much has been made in recent years about the effects of orientation on the discovery and classification of active galaxies. Antonucci (1993) stresses polarimetric evidence that many AGN contain opaque tori and that consideration of the effects of such structures can be taken to "unify" what have traditionally been considered distinct classes of objects. NGC 4258 has provided a key element to that argument by revealing a direct link between the orientations of an inferred obscuring torus and a maser disk which orbits a compact object of very large mass. The fact that this has been deduced from an object which has been termed a LINER implies not only that the unification picture can be applied to AGN over several decades in luminosity (cf. Hines et al. 1995), but also that all AGN are powered, in part, by accretion of material onto a central massive object.

GDS thanks the staff of Lowell Observatory for their hospitality and telescope time during a short sabbatical leave. BJW thanks Oxford University Astrophysics Department for their hospitality during the completion of this work. We would also like to thank the referee, Dean Hines, for his careful comments which have added to the clarity of the paper.

Financial support was provided by grants NSF AST 91–14087 (GDS), NASA NAG 5–1630 (PSS), NASA NAG W–3134 (KKM, BJW), and NAG W5–2201, NAG W–4490 (SM).

Table 1. Measurements of Polarized Emission Lines and the Continuum

| Line | $\lambda_{obs}$ (Å) | Stokes Flux[a] ($\times 10^{-15}$) | FWHM (km s$^{-1}$) | $p$ (%) | PA (°) |
|---|---|---|---|---|---|
| H$\beta$ | 4871.1 | 0.92: | 1100 | ... | ... |
| [OIII] $\lambda$4959 | 4966.1 | 0.85 | 800 | 1.1 ±0.4 | 84± 9 |
| [OIII] $\lambda$5007 | 5014.6 | 3.10 | 1300 | 3.0 ±0.3 | 83± 3 |
| [OI] $\lambda$6300 | 6310.6 | 0.46: | 310 | 4.6 ±1.8 | 77±11 |
| [OI] $\lambda$6363 | 6374.3 | 0.27: | 610: | ... | ... |
| [NII] $\lambda$6548 | 6557.9 | 0.75: | 1030: | 3.4 ±1.0 | 85± 8 |
| H$\alpha$ | 6576.8 | 4.27 | 980 | 9.4 ±0.5 | 88± 2 |
| [NII] $\lambda$6583 | 6593.3 | 2.31 | 870 | 4.5 ±0.7 | 87± 5 |
| [SII] $\lambda\lambda$6716 + 6731 | 6742.1 | 0.59 | ... | 1.0 ±0.5 | 77±14 |
| Continuum (slit EW) | 5100 − 6500 | | | 0.23±0.02 | 83± 3 |
| Continuum (slit NS) | 5100 − 6500 | | | 0.23±0.01 | 79± 2 |

[a]$q' \times F$ in ergs cm$^{-2}$ s$^{-1}$ for a coordinate system rotated to the systemic position angle (PA) of polarization (here 85°). Stokes flux is equivalent to polarized flux but avoids the bias and peculiar error distribution associated with $p = \sqrt{q^2 + u^2}$. Estimates for individual lines in the H$\alpha$/[NII] complex required deblending.

## Figure Captions

Figure 1: Polarization and surface brightness map of the central 47″×47″ (1.6×1.6 kpc) of NGC 4258 obtained in the photometric $V$-band. Contours represent 20% increments in total surface brightness, while the surface brightness and position angle of polarized light is shown on a 1″ (35 pc) grid by short line segments. Note the centrally concentrated peak in polarized flux with an EW position angle, identical to the plane of the $H_2O$ maser disk. Several contour intervals in the central brightness peak have been omitted for clarity.

Figure 2: Nuclear spectra of NGC 4258 in total flux (*top*) and Stokes flux (*bottom*; see also Table 1). In Stokes flux, a prominent emission-line spectrum with FWHM ∼1000 km s$^{-1}$ is superposed on a continuum of approximate shape $F_\nu \propto \nu^{-1.1}$.

Figure 3: The observed (open boxes), polarized and inferred nuclear continuum levels of NGC 4258 compared with a typical radio-quiet quasar (RQQ) spectral energy distribution and the ionizing continuum from Stüwe *et al.* (1992, dashed line). The optical and near infrared data, which seems to be dominated by the galactic bulge, were obtained on the SAO 1.22 m and Steward Observatory 1.55 m telescopes in June 1995. Other references are: UV: Ellis, Gondhalekar, & Efstathiou (1982); X-ray: Makishima *et al.* (1994); radio: Hummel *et al.* (1989); Turner & Ho (1994); and IR: Rieke & Lebofsky (1978); Dyck, Becklin, & Capps (1978).

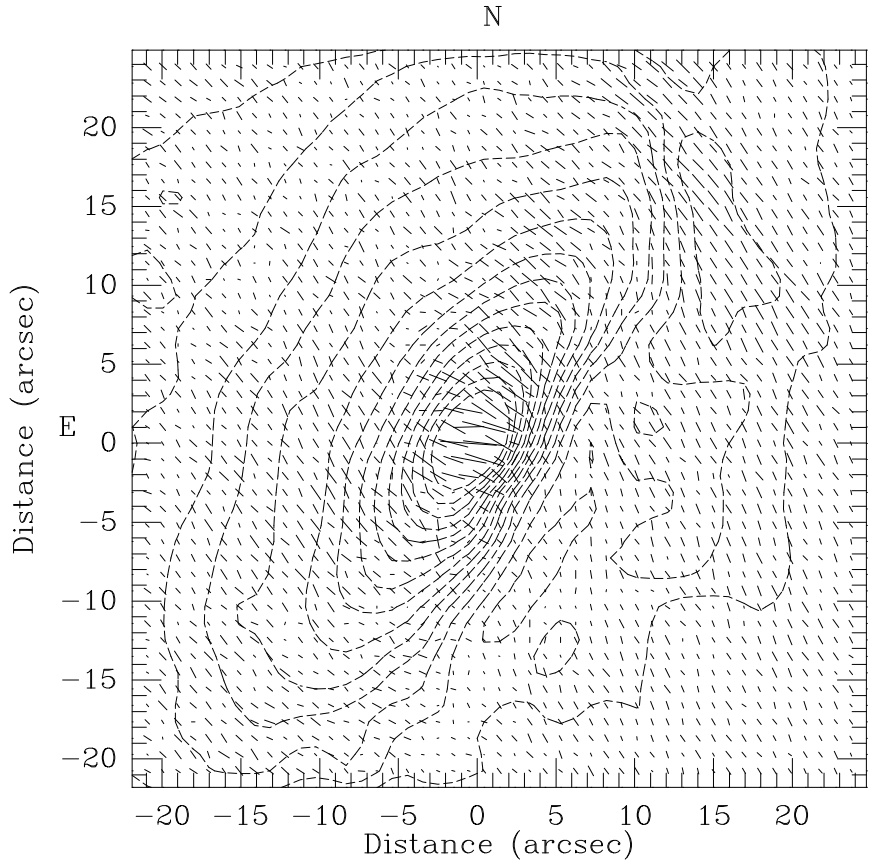

Fig. 1.— Polarization and surface brightness map of the central $47'' \times 47''$ ($1.6 \times 1.6$ kpc) of NGC 4258 obtained in the photometric $V$-band. Contours represent 20% increments in total surface brightness, while the surface brightness and position angle of polarized light is shown on a $1''$ (35 pc) grid by short line segments. Note the centrally concentrated peak in polarized flux with an EW position angle, identical to the plane of the $H_2O$ maser disk. Several contour intervals in the central brightness peak have been omitted for clarity.

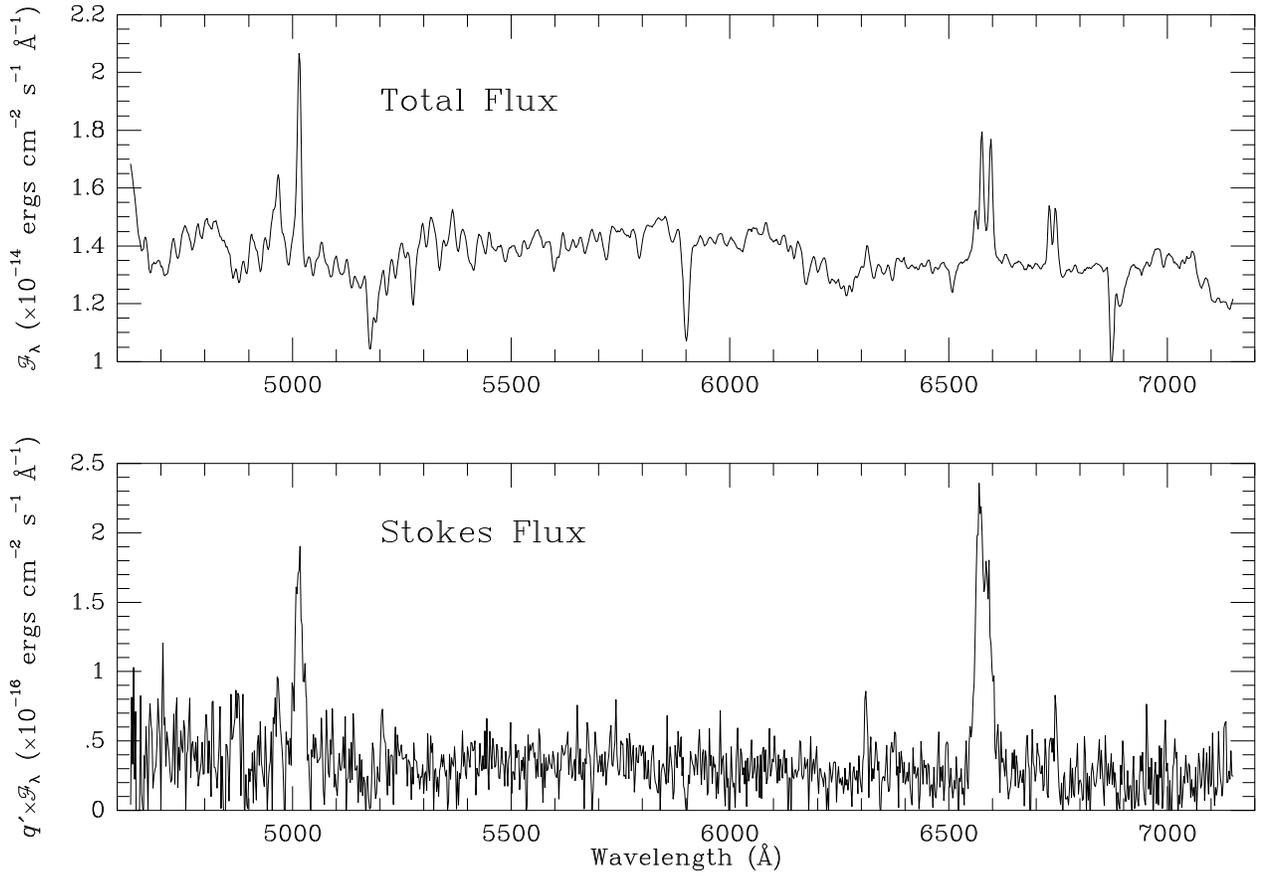

Fig. 2.— Nuclear spectra of NGC 4258 in total flux (*top*) and Stokes flux (*bottom*; see also Table 1). In Stokes flux, a prominent emission-line spectrum with FWHM ∼1000 km s$^{-1}$ is superposed on a continuum of approximate shape $F_\nu \propto \nu^{-1.1}$.

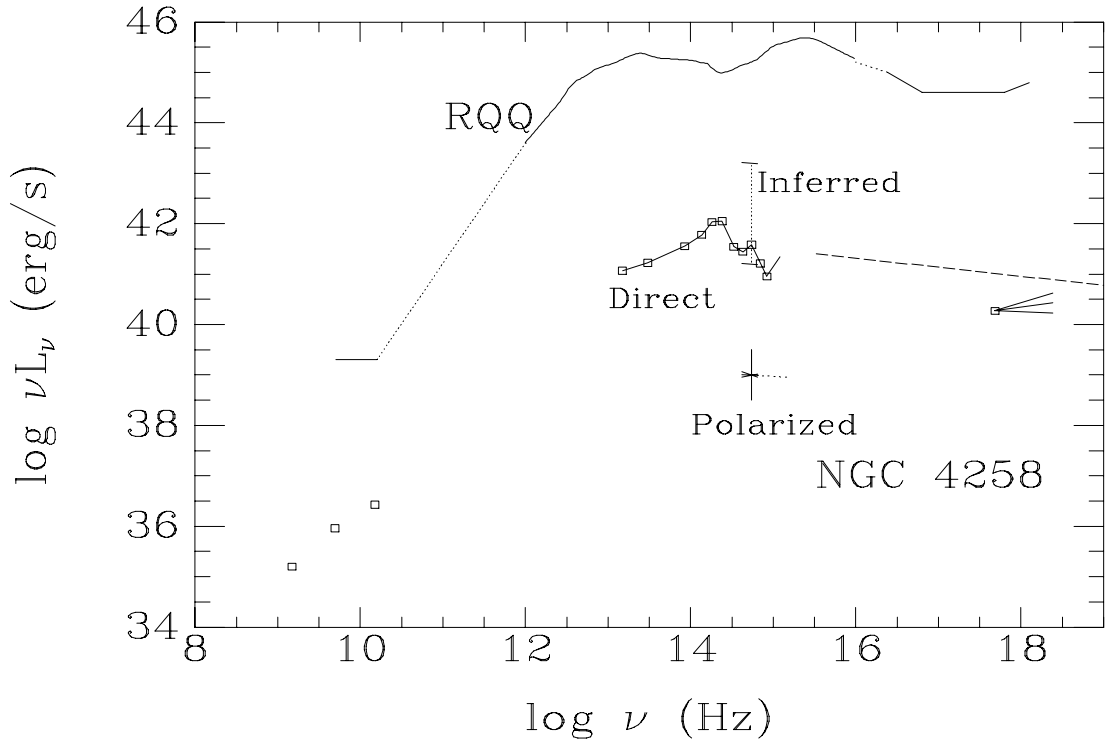

Fig. 3.— The observed (open boxes), polarized and inferred nuclear continuum levels of NGC 4258 compared with a typical radio-quiet quasar (RQQ) spectral energy distribution and the ionizing continuum from Stüwe *et al.* (1992, dashed line). The optical and near infrared data, which seems to be dominated by the galactic bulge, were obtained on the SAO 1.22 m and Steward Observatory 1.55 m telescopes in June 1995. Other references are: UV: Ellis, Gondhalekar, & Efstathiou (1982); X-ray: Makishima *et al.* (1994); radio: Hummel *et al.* (1989); Turner & Ho (1994); and IR: Rieke & Lebofsky (1978); Dyck, Becklin, & Capps (1978).